**Title Page**

**Title:** A No-Go Framework for Coarse-Grained Effect Prediction in Boron Neutron Capture Therapy: Structural Identifiability Limits and Microscopic Coverage

**Authors**: Atsushi Fujimura[1*], Shuichi Furuya[2], Yasuaki Ichikawa[2]

[1] Department of Molecular Physiology, Kagawa University Faculty of Medicine, 1750-1 Ikenobe, Miki-cho, Kita-gun, Kagawa 761-0793, Japan.

[2] Particle Beam Therapy Research Institute, Inc., c/o Tanaka Law Office, Kanda Nishiki-cho 1-21-1, Chiyoda-ku, Tokyo 101-0054, Japan.

* corresponding author

**Corresponding author:**

Atsushi Fujimura, M.D., Ph.D.

Department of Molecular Physiology, Kagawa University Faculty of Medicine, 1750-1 Ikenobe, Miki-cho, Kita-gun, Kagawa 761-0793, Japan.

Tel: +81-87-891-2095

e-mail: fujimura.atsushi@kagawa-u.ac.jp

**Abstract**

**Objective.** To establish when coarse-grained observables uniquely determine an expectation-based biological response in boron neutron capture therapy (BNCT), providing a structural identifiability analysis of the microscopic information hierarchy underlying effect prediction.

**Approach.** We represent microscopic target populations by probability distributions and retained observables by finitely many population-averaged linear descriptors. We derive an exact-identification criterion, ambiguity and approximation bounds, and partial-identification intervals. A minimal marked-Poisson capture-event model links the framework to BNCT, while additional survival-kernel and temporal assumptions support conditional biological results.

**Main results.** For an admissible class containing all finitely supported microscopic distributions, exact identification holds if and only if the response kernel is a constant plus a linear combination of the retained descriptor functions. Mean dose, mixed-field component means, and finite moment sets are therefore not generally sufficient for exact identification over unrestricted microscopic heterogeneity. Bounded, non-increasing convex survival kernels yield sharp survival intervals given a mean and an upper support bound and a treatment-ranking criterion. Lower-tail bounds and high-exposure limits require bounded, non-increasing survival kernels; contributions from relative exposure bounded away from zero vanish when the kernel tends to zero. Conditional temporal re-sampling at fixed total mean exposure reduces transient heterogeneity while preserving differences in persistent target-specific mean exposure. In the marked-event specialization, separate frequency and quality distributions need not determine survival.

**Significance.** Failure of exact identification does not preclude useful approximation, partial identification, or model-dependent prediction. The framework identifies information lost through BNCT coarse-graining and clarifies the relevance of microscopic lower-tail, longitudinal, and joint exposure–biological context information, without prescribing a measurement modality.

## 1. Introduction

Boron neutron capture therapy (BNCT) is intrinsically multiscale. Neutron transport determines the radiation field at the tissue level, $^{10}B$ pharmacokinetics determines the availability of capture targets, individual $^{10}B(n,\alpha)^{7}Li$ reactions generate short-range high-linear-energy-transfer charged particles, and their biological consequences depend on the microscopic spatial and cellular context in which those reactions occur (Coderre and Morris 1999; International Atomic Energy Agency 2001; Sauerwein et al 2025). The development and clinical translation of BNCT have therefore required close integration of neutron physics, radiation biology, boron chemistry, drug delivery, imaging, dosimetry, and oncology (Barth et al 2012; Jin et al 2022; Malouff et al 2021; Moss 2014).

Selective boron delivery and the ability to measure boron concentration and spatial distribution have long been recognized as central requirements for BNCT (Barth et al 2018; Wittig et al 2008). Clinical and experimental dosimetry necessarily compresses this microscopic complexity into a limited number of quantitative descriptors. Physical dose is conventionally decomposed into boron, nitrogen, hydrogen or fast-neutron, and photon components, which may subsequently be combined using biological weighting factors or incorporated into more elaborate radiobiological frameworks (Coderre and Morris 1999; González and Santa Cruz 2012; International Atomic Energy Agency 2001; Sauerwein et al 2025). Such quantities are indispensable for treatment planning, prescription, and comparison. Their practical usefulness, however, does not imply that they retain all microscopic information relevant to biological effect.

The importance of microscopic heterogeneity has long been recognized. Gabel and colleagues showed that $^{10}B$ capture reactions generate broad distributions of nuclear hit size and emphasized that boron localization relative to radiation-sensitive cellular structures strongly modifies biological effect (Gabel et al 1987). Subsequent stochastic microdosimetric approaches explicitly incorporated intra- and intercellular heterogeneity in $^{10}B$ distribution into effect prediction (Sato et al 2018). Cellular-scale Particle and Heavy Ion Transport code System (PHITS) simulations further demonstrated that intracellular boron localization can markedly alter the biological effect associated with a given amount of boron (Shigehira et al 2023), and the subsequent development of Linear Interpolation System for SMK Model Parameters Evaluated from Cellular-Scale Simulation (LISMEC) enabled cellular-scale stochastic microdosimetric kinetic (SMK) parameters to be propagated efficiently into BNCT dosimetry (Shigehira et al 2026). More recent microdosimetric analyses have reinforced the importance of small-number capture-event statistics and experimentally informed subcellular boron microdistribution (Marcaccio et al 2026; Santa Cruz 2026).

In parallel, increasingly detailed experimental and imaging studies have revealed heterogeneity across spatial and cellular scales. Treatment-planning analyses incorporating heterogeneous boron distributions have shown that uniform-uptake assumptions can misrepresent low-uptake regions (Teng et al 2023), whereas spatial positron emission tomography (PET) analyses have emphasized the importance of non-mean measures of intratumoral uptake (Nakaichi et al 2022). More recently, single-cell mass-spectrometry imaging has directly demonstrated striking cell-to-cell heterogeneity in boronophenylalanine (BPA) uptake within experimental glioblastoma tumors (Fang et al 2025). Together, these observations establish heterogeneity across spatial and cellular scales as a practical challenge for contemporary BNCT (Zhou et al 2025).

These experimental and modeling advances show that microscopic heterogeneity can materially affect BNCT response prediction (Gabel et al 1987; Marcaccio et al 2026; Santa Cruz 2026; Sato et al 2018; Shigehira et al 2023; Shigehira et al 2026). They do not, however, resolve a logically distinct question: whether the information required by such models is itself recoverable from the coarse-grained observables available for prediction. A forward model asks what biological effect follows once the relevant microscopic state, distribution, or radiobiological structure has been specified. An identifiability analysis instead asks whether that response is uniquely determined by the observables retained after coarse-graining. A forward model may therefore be internally well defined while its biological prediction remains non-identifiable from an insufficient observational representation.

This distinction is particularly relevant to BNCT because averaging occurs at several levels. Mean boron concentration does not specify its cell-to-cell distribution. Mean physical dose does not specify the microscopic distribution of capture events or energy deposition. Separate averages of mixed-field components do not generally specify their joint exposure within the same biological targets. Likewise, measurements of boron quantity, spatial distribution, and biological state obtained separately need not determine how these properties co-occur within individual clonogens. Advances in boron analysis and imaging (Wittig et al 2008), together with modern spatial and single-cell approaches (Fang et al 2025; Marcaccio et al 2026; Nakaichi et al 2022; Teng et al 2023), increasingly provide access to such heterogeneity, but they also raise a more fundamental question:

*What biological information is mathematically recoverable from the coarse-grained observables retained for prediction?*

In a recent conceptual study, we organized BNCT effect prediction into a hierarchy linking observable information, latent cellular boron state, cell-associated capture statistics, microdosimetric context, and biological consequence (Furuya and Fujimura 2026). That framework distinguished capture frequency from event-relevant microscopic context, but did not address whether the information retained after coarse-graining is sufficient to determine biological response uniquely. The present study formalizes that question as one of structural identifiability. The transition from multiscale BNCT information to finite-summary structural identifiability is summarized in figure 1.

We represent a heterogeneous microscopic target population by a probability distribution and its coarse-grained representation by finitely many population summaries. We then ask when an expectation-based biological response is uniquely determined by those summaries without imposing a parametric form on the underlying microscopic distribution.

We first derive a necessary and sufficient criterion for exact identification from finitely many linear population summaries and apply it to mean dose, mixed-field component summaries, finite moments, and capture-event statistics. We then distinguish exact identification from partial identification and controlled approximation. Under additional assumptions on the survival kernel, we derive sharp response bounds, a treatment-ranking criterion, and lower-tail constraints, followed by extensions to temporal re-sampling and joint dependence between reaction burden and event-relevant biological context.

In this paper, microscopic coverage refers to information about the lower-exposure fraction of the target population, rather than to a new dosimetric quantity. The contribution of this study is to translate microscopic coarse-graining in BNCT into an explicit structural-identifiability problem and to determine when finite summaries preserve enough information for exact recovery. The framework complements macroscopic dosimetry and radiobiological forward models: those models address how a specified microscopic state maps to biological effect, whereas identifiability analysis asks whether the retained observables determine that response uniquely.

Here, the term No-Go refers specifically to failure of unique exact recovery under the stated information constraints; it does not imply that useful prediction is impossible. When exact identification fails, partial identification, controlled approximation, or prediction under independently justified model restrictions may remain possible. The biological consequences developed below are conditional on the stated response-kernel and exposure-model assumptions; optimal measurement selection and budget-constrained design are outside the present scope.

## 2. Results

### 2.1. Structural identifiability under microscopic coarse-graining

#### 2.1.1. Microscopic states, biological response, and finite summaries

BNCT effect prediction requires a heterogeneous microscopic target population to be represented by a finite set of macroscopic observables. To formalize the information lost in this reduction, let $X$ denote the space of physically and biologically admissible microscopic target states, equipped with a $\sigma$-algebra $\Sigma$. A state $x \in X$ may include local boron availability, neutron exposure, subcellular localization, cellular geometry, intrinsic susceptibility, or any other target-specific variable relevant to response.

Let $P$ be a probability measure describing the distribution of microscopic states across the biological target population. Any fixed biological weighting of targets can be incorporated into the definition of $P$.

Let $g: X \to \mathbb{R}$ be a measurable single-target response kernel. For an expectation-based endpoint, the corresponding population response is

$$R_g(P) = \int_X g(x)\, dP(x). \tag{1}$$

For clonogenic survival, $0 \leq g(x) \leq 1$. Unless otherwise stated, all response kernels and descriptor functions are assumed measurable, and all expectations appearing in a stated result are assumed finite. Results involving moments additionally require the existence of the corresponding moments; when bounded support is assumed, the required moment conditions are automatically satisfied.

We next represent the information retained after microscopic coarse-graining. Let $h_j: X \to \mathbb{R}$, $j = 1, \ldots, m$, denote measurable microscopic descriptor functions. Their population averages are

$$m_j(P) = \int_X h_j(x)\, dP(x), \tag{2}$$

and the complete retained summary vector is

$$T_H(P) = \big(m_1(P), \ldots, m_m(P)\big). \tag{3}$$

The reduction from a heterogeneous microscopic population to a finite retained representation is illustrated in figure 1(b).

The term *linear summary* refers here to linearity with respect to the population distribution $P$, rather than to linearity of $h_j(x)$ in the microscopic state. Means, raw moments, fractions of targets satisfying specified conditions, and histogram-bin occupancies all belong to this class.

For example, variance is not itself linear in $P$, but retaining the first two raw moments is informationally equivalent to retaining the mean and variance because

$$\mathrm{Var}(\Lambda) = E[\Lambda^2] - E[\Lambda]^2. \tag{4}$$

We use exact identification to mean that the retained representation uniquely determines the population response; structural identifiability refers to this property over the specified admissible class.

**Definition 1. Exact distribution-free structural identifiability.** The population response $R_g$ is said to be *exactly distribution-free identifiable* from $T_H$ if, for every pair of admissible microscopic distributions $P_1$ and $P_2$,

$$T_H(P_1) = T_H(P_2) \quad \Rightarrow \quad R_g(P_1) = R_g(P_2). \tag{5}$$

Equivalently, there exists some function $F$, not necessarily linear, such that

$$R_g(P) = F\big(T_H(P)\big) \tag{6}$$

for every admissible $P$.

Thus, exact identifiability permits arbitrarily nonlinear post-processing. The issue is whether the retained observables contain enough information to determine the response uniquely.

### 2.1.2. A finite-summary identifiability theorem

Let

$$V = \mathrm{span}\{1, h_1, \dots, h_m\} \tag{7}$$

Here, the constant function is included because probability measures are normalized. Membership in this space therefore means that the response kernel is a constant plus a linear combination of the retained descriptor functions.

Lemma 1 supplies the finite evaluation set used in the necessity proof.

**Lemma 1. Finite evaluation lemma.** If $f_1, \dots, f_n$ are linearly independent real-valued functions on $X$, then there exist $x_1, \dots, x_n \in X$ such that the evaluation matrix

$$M_{ij} = f_i(x_j) \tag{8}$$

is nonsingular.

*Proof.* The claim is immediate for $n = 1$. Assume that it holds for $n - 1$ linearly independent functions, and choose $x_1, \dots, x_{n-1}$ such that their $(n-1) \times (n-1)$ evaluation matrix is nonsingular.

Now include $f_n$ and allow the final evaluation point $x$ to vary over $X$. If the determinant of the resulting $n \times n$ matrix were zero for every $x \in X$, expansion along the final column would express $f_n(x)$ as a fixed linear combination of $f_1(x), \dots, f_{n-1}(x)$ for every $x$, contradicting linear independence.

Therefore, at least one point $x_n$ exists for which the complete evaluation matrix is nonsingular. ■

**Theorem 1. Finite linear-summary identifiability theorem.** Suppose that the admissible class of microscopic distributions contains all finitely supported probability measures on $X$. Then

$$R_g(P) = \int_X g(x)\, dP(x) \tag{9}$$

is exactly distribution-free identifiable from $T_H(P)$ if and only if

$$g \in V. \tag{10}$$

Equivalently, exact identification is possible if and only if constants $a_0, a_1, \dots, a_m$ exist such that

$$g(x) = a_0 + \sum_{j=1}^{m} a_j\, h_j(x) \tag{11}$$

for every $x \in X$.

*Proof. Sufficiency.* If the representation above holds, then

$$R_g(P) = a_0 + \sum_{j=1}^{m} a_j \, m_j(P), \tag{12}$$

and the population response is determined uniquely by $T_H(P)$.

*Necessity.* Suppose instead that $g \notin V$. Let $v_1, \ldots, v_d$ be a basis of $V$, chosen such that $v_1(x) = 1$, where $d = \dim V$.

Because $g \notin V$, the functions $v_1, \ldots, v_d, g$ are linearly independent. By Lemma 1, there exist $x_1, \ldots, x_{d+1} \in X$ such that their complete $(d+1) \times (d+1)$ evaluation matrix is nonsingular.

Let $A$ denote the $d \times (d+1)$ submatrix containing only the rows corresponding to $v_1, \ldots, v_d$. Since $\operatorname{rank}(A) = d$, there exists a non-zero vector

$$a = (a_1, \ldots, a_{d+1}) \tag{13}$$

such that

$$Aa = 0. \tag{14}$$

Hence,

$$\sum_{j=1}^{d+1} a_j \, v_k(x_j) = 0 \tag{15}$$

for every $k = 1, \ldots, d$. Because $v_1(x) = 1$,

$$\sum_{j=1}^{d+1} a_j = 0. \tag{16}$$

The nonsingularity of the complete evaluation matrix further implies

$$\sum_{j=1}^{d+1} a_j \, g(x_j) \neq 0. \tag{17}$$

The coefficients $a_j$ must therefore contain both positive and negative values. Let $J_p$ denote the indices for which $a_j > 0$, and let $J_n$ denote the indices for which $a_j < 0$.

Define

$$s_a = \sum_{j \in J_p} a_j. \tag{18}$$

Because $\sum_{j=1}^{d+1} a_j = 0$, the same positive scalar also satisfies

$$s_a = -\sum_{j\in J_n} a_j > 0. \tag{19}$$

Construct two probability measures

$$P_a = \sum_{j\in J_p} \frac{a_j}{s_a} \delta_{x_j}, \qquad P_b = \sum_{j\in J_n} \frac{-a_j}{s_a} \delta_{x_j}. \tag{20}$$

For every function in $V$, these two distributions have identical expectations. Therefore,

$$T_H(P_a) = T_H(P_b), \tag{21}$$

whereas

$$R_g(P_a) - R_g(P_b) = \frac{1}{s_a}\sum_{j=1}^{d+1} a_j\, g(x_j) \neq 0. \tag{22}$$

Thus, exact identifiability fails whenever $g \notin V$. ∎

Theorem 1 therefore characterizes exactly when an expectation-based response can be recovered from a prescribed finite set of linear population summaries.

This statement is distinct from Jensen's inequality. Theorem 1 concerns exact recovery and requires no convexity assumption; Jensen's inequality provides orderings or bounds only after a convex response kernel and an appropriate summary have been specified.

The necessity construction uses only finitely supported populations, so non-identifiability does not depend on pathological continuous distributions. Two finite microscopic populations can share all retained summaries while producing different responses, as illustrated in figure 1(c); the corresponding exact-identification criterion is summarized in figure 1(d).

Finally, because the mapping $F$ was allowed to be arbitrarily nonlinear, nonlinear computation applied after coarse-graining cannot recover microscopic information that was not retained before the prediction step.

### 2.1.3. Identified sets, irreducible ambiguity, and restricted model families

For a given attainable summary vector $t$, define the *identified response set*

$$I_g(t) = \{R_g(P): T_H(P) = t\}. \tag{23}$$

Exact identifiability at $t$ is equivalent to $I_g(t)$ containing a single value.

Suppose instead that two microscopic distributions $P_a$ and $P_b$ satisfy

$$T_H(P_a) = T_H(P_b) = t \tag{24}$$

but $R_g(P_a) \neq R_g(P_b)$. Any prediction rule using only the retained summaries must assign the same predicted response $F(t)$ to both populations.

**Corollary 1. Irreducible ambiguity bound.** For every prediction rule $F$ based solely on $t$, at least one of the two errors $\left|F(t) - R_g(P_a)\right|$ and $\left|F(t) - R_g(P_b)\right|$ is no smaller than one-half of the response separation $\left|R_g(P_a) - R_g(P_b)\right|$.

*Proof.* By the triangle inequality,

$$\left|R_g(P_a) - R_g(P_b)\right| \leq \left|F(t) - R_g(P_a)\right| + \left|F(t) - R_g(P_b)\right|. \tag{25}$$

The stated bound follows immediately. ■

When a retained summary is compatible with more than one response, the identified set records that ambiguity. Corollary 1 then gives an irreducible worst-case error bound for any point predictor restricted to the same summaries.

The present analysis concerns structural identifiability in the noiseless population limit. The summaries $T_H(P)$ are assumed to be known exactly. Finite sampling, imaging noise, measurement uncertainty, reconstruction error, and statistical estimation constitute separate problems.

The ambiguity is therefore structural rather than observational.

The necessity statement also depends on the admissible class of microscopic distributions. If independent physical or biological information restricts $P$ to a smaller family $\mathcal{P}_0$, finite summaries may become sufficient even when $g \notin V$.

Such a result would not contradict Theorem 1: the restriction

$$P \in \mathcal{P}_0 \tag{26}$$

is itself additional information. Distribution-free structural identifiability must therefore be distinguished from identification conditional on a restricted parametric or mechanistic family, whose validity depends on independent physical or biological justification.

Exact identification, partial identification, controlled approximation, and model-dependent prediction are distinct tasks. Failure of exact identification does not preclude useful results from the latter three.

#### 2.1.4. Consequences for mean and mixed-field dose

We first consider a target-weighted mean boron-dose descriptor under matched physical and biological weighting.

Let $d_B(x)$ denote a microscopic boron-dose-related descriptor defined for the same biological targets represented by $P$. Define the corresponding target-weighted mean boron dose as

$$D_{B,\text{mean}} = \int_X d_B\,(x)\,dP(x). \tag{27}$$

**Corollary 2. Mean boron-dose identifiability criterion.** The biological response $R_g(P)$ is exactly distribution-free identifiable from $D_{B,\text{mean}}$ alone if and only if

$$g(x) = a + b\,d_B(x) \tag{28}$$

for constants $a$ and $b$.

Corollary 2 shows that, even under matched weighting, the mean is exactly sufficient only when the microscopic response is affine in the averaged dose descriptor.

If this condition fails on the admissible state space, distinct microscopic populations can share the same mean while producing different responses.

Any mismatch between clinical voxel-, volume-, or mass-weighted dose and clonogen-weighted exposure is a separate layer of information loss; it is not required for the result above.

BNCT is also a mixed radiation field. Let $C$ denote a finite set of radiation components, and let $d_c(x)$ denote the microscopic target-associated descriptor for component $c \in C$. Define

$$D_{c,\text{mean}} = E[d_c(x)]. \tag{29}$$

**Corollary 3. Mixed-field component-mean criterion.** The complete vector of component means exactly identifies $R_g$ if and only if

$$g \in \text{span}\{1, d_c : c \in C\}. \tag{30}$$

For example, separate knowledge of $E[d_B]$ and $E[d_\gamma]$ does not, in general, determine

$$E[d_B d_\gamma]. \tag{31}$$

Thus, joint component information is required whenever the complete response kernel lies outside the span generated by the separate component descriptors in equation (30).

The same criterion applies to a fixed weighted dose

$$D_w = \sum_{c \in C} w_c \, D_{c,\text{mean}}, \tag{32}$$

where

$$d_w(x) = \sum_{c \in C} w_c \, d_c(x), \tag{33}$$

exact identification from $D_w$ alone requires

$$g(x) = a + b \, d_w(x). \tag{34}$$

These statements concern information sufficiency rather than the validity of nonlinear effect-equivalent or microdosimetric forward models. Such models may supply additional physical or biological structure, but their predictions still depend on the information provided to them.

### 2.1.5. A minimal $^{10}B$ capture-event specialization

To connect the abstract criterion to BNCT capture physics, we next define a target-associated reaction-intensity descriptor.

For target $i$, let $\Omega_i$ denote a specified reaction-scoring domain. Let $n_{10B}(\mathbf{r}, t)$ be the local $^{10}$B number density, $\phi_n(\mathbf{r}, E, t)$ the differential scalar neutron flux, and $\sigma_{10B}(E)$ the microscopic cross section for the $^{10}B(n,\alpha)^{7}Li$ reaction.

The expected number of target-associated physical capture reactions during irradiation from time 0 to $T$ is

$$\lambda_i = \int_0^T \int_{\Omega_i} \int_0^\infty n_{10B}(\mathbf{r}, t)\, \phi_n(\mathbf{r}, E, t)\, \sigma_{10B}(E)\, dE\, dV\, dt. \quad (35)$$

The corresponding dimensions are

$$[n_{10B}] = L^{-3}, \qquad [\phi_n] = L^{-2}T^{-1}\mathcal{E}^{-1}, \qquad [\sigma_{10B}] = L^2, \quad (36)$$

where $L$, $T$, and $\mathcal{E}$ denote the dimensions of length, time, and energy, respectively; after integration over energy, volume, and time, $\lambda_i$ is dimensionless.

The scoring domains $\Omega_i$ need not form a partition of tissue and may overlap. Accordingly, $\lambda_i$ should be interpreted as a target-associated expected reaction count under the stated scoring convention, rather than as an energy-conserving partition of physical reactions or imparted energy among cells; nor is it itself an ICRU microdosimetric quantity such as specific energy $z$ or lineal energy $y$ (International Commission on Radiation Units and Measurements 1983).

Sampling $\lambda_i$ across the target population defines a non-negative random variable $\Lambda$, with

$$\mu = E[\Lambda]. \quad (37)$$

For a specified neutron field, reference geometry, and scoring convention, a nominal target-population mean boron-dose scale may be written

$$D_{B,\text{mean}}^{(\text{nom})} = \kappa\mu, \quad (38)$$

where $\kappa$ has units of Gy per expected reaction under that scoring convention. This proportionality is not required for the general identifiability theorem.

For an analytically transparent nonlinear specialization, consider the conditional event-count model

$$K \mid \Lambda = \lambda \sim \text{Poisson}(\lambda), \quad (39)$$

where $K$ is the realized number of target-associated capture events.

Let $R_j \in [0,1]$ denote the multiplicative survival contribution associated with event $j$. Conditional on $(K = k, \Lambda = \lambda)$, assume that the $R_j$ are independent and identically distributed with common conditional mean $r$, independent of $j$, $k$, and $\lambda$.

Define

$$q = 1 - r. \tag{40}$$

Then

$$E\left[\prod_{j=1}^{k} R_j \mid K = k, \Lambda = \lambda\right] = r^k, \tag{41}$$

Averaging over the Poisson event count gives

$$g_q(\lambda) = \sum_{k=0}^{\infty} e^{-\lambda} \frac{\lambda^k}{k!} r^k = e^{-q\lambda}. \tag{42}$$

Thus, the corresponding population survival is

$$S_{\exp}(q) = E[e^{-q\Lambda}]. \tag{43}$$

The subscript 'exp' distinguishes this exponential specialization from the general survival kernels considered below.

The parameter $q$ is an abstract effective event-quality parameter. The marked-Poisson assumptions provide an analytically transparent connection between heterogeneous capture-event intensity and a nonlinear response; their biological scope is discussed in Section 3.2.

### 2.1.6. Finite summaries, approximation, and partial identification

For $q > 0$, the function

$$g_q(\lambda) = e^{-q\lambda} \tag{44}$$

is nonlinear. By Theorem 1, the population response $E[e^{-q\Lambda}]$ is therefore not exactly distribution-free identifiable from $E[\Lambda]$ alone.

For a positive common mean reaction intensity, compare a homogeneous population A,

$$P_A(\Lambda = \mu) = 1, \tag{45}$$

with a two-point heterogeneous population B,

$$P_B(\Lambda = 0) = 1 - p, \qquad P_B(\Lambda = \mu/p) = p, \tag{46}$$

with $0 < p < 1$.

Both populations have the same mean,

$$E_A[\Lambda] = E_B[\Lambda] = \mu, \tag{47}$$

but different survival values:

$$S_A = e^{-q\mu}, \qquad S_B = 1 - p + p\, e^{-q\mu/p}. \tag{48}$$

Strict convexity of $e^{-q\lambda}$ gives

$$S_B > S_A \tag{49}$$

for $0 < p < 1$. Furthermore, $S_B$ approaches 1 as $p$ approaches zero from above.

Thus, the same positive mean reaction intensity can correspond to substantially different biological responses. This same-mean/different-survival construction is illustrated in figure 2(a).

The example is structural rather than representative: excluding such distributions requires information beyond the mean.

The same principle extends to finite moment descriptions. Suppose the first $n$ raw moments of $\Lambda$ are retained. Their descriptor space is

$$\mathrm{span}\{1, \lambda, \lambda^2, \dots, \lambda^n\}. \tag{50}$$

**Corollary 4. Finite-moment criterion.** The expectation $E[g(\Lambda)]$ is exactly distribution-free identifiable from the first $n$ moments of $\Lambda$ if and only if $g$ coincides on the admissible state space with a polynomial of degree at most $n$.

For the exponential kernel, a polynomial of degree $n$ cannot coincide with $e^{-q\lambda}$ at $n + 2$ or more distinct values of $\lambda$. Therefore, in the usual continuous or otherwise infinite-state setting, no finite collection of reaction-intensity moments exactly identifies $E[e^{-q\Lambda}]$.

Exact non-identifiability does not, however, imply poor approximation.

Let

$$v(x) = a_0 + \sum_{j=1}^{m} a_j\, h_j(x) \tag{51}$$

belong to $V$. Let $\varepsilon_v$ denote the supremum of $|g(x) - v(x)|$ over $x \in X$, whenever this supremum is finite.

**Proposition 1. Summary-compatible approximation bound.** For every admissible probability distribution $P$,

$$\left|R_g(P) - R_v(P)\right| \le \varepsilon_v. \tag{52}$$

*Proof. The triangle inequality gives*

$$\left|R_g(P) - R_v(P)\right| \le \int_X |g(x) - v(x)|\; dP(x) \le \varepsilon_v. \tag{53}$$

■

Thus, a reduced representation can remain uniformly accurate when the true response kernel is uniformly close to the retained linear span, even if exact identification fails.

For the exponential specialization, additionally assume bounded support,

$$0 \le \Lambda \le \Lambda_{\max} \tag{54}$$

and approximate the response kernel by the degree-$n$ Taylor polynomial. The resulting population approximation is

$$S_n(q) = \sum_{k=0}^{n} \frac{(-q)^k}{k!} E[\Lambda^k]. \tag{55}$$

**Proposition 2. Finite-moment approximation bound.**

$$\left|S_{\exp}(q) - S_n(q)\right| \le \frac{q^{n+1} E[\Lambda^{n+1}]}{(n+1)!} \le \frac{(q\Lambda_{\max})^{n+1}}{(n+1)!}. \tag{56}$$

*Proof.* For $0 \le \lambda \le \Lambda_{\max}$, Taylor's theorem gives

$$\left|e^{-q\lambda} - \sum_{k=0}^{n} \frac{(-q\lambda)^k}{k!}\right| \le \frac{(q\lambda)^{n+1}}{(n+1)!}, \tag{57}$$

because the absolute value of the $(n+1)$-th derivative of $e^{-q\lambda}$ is bounded by $q^{n+1}$. Taking expectations gives the first inequality, and

$$E[\Lambda^{n+1}] \leq \Lambda_{\max}^{n+1} \tag{58}$$

gives the second. ■

Together, Propositions 1 and 2 separate exact sufficiency from controlled approximation: exact identification may fail while finite summaries still support uniform error bounds under additional regularity or support constraints.

We next ask what response statements remain available when the microscopic exposure distribution is only partially constrained.

## 2.2. Conditional biological consequences of unresolved microscopic heterogeneity

Theorem 1 itself does not require convexity. The results in this section add explicit structure: convexity for mean-and-support and convex-order results; bounded monotonicity for lower-tail and high-exposure results; and additional model assumptions for TCP propagation, temporal re-sampling, and event-quality coupling.

### 2.2.1. Survival bounds under convexity assumptions

To determine which consequences are specific to the exponential specialization, let $s$ be a bounded, non-increasing, convex single-target survival kernel defined for non-negative exposure values, and define

$$S_s = E[s(\Lambda)]. \tag{59}$$

As a survival probability, the kernel takes values between zero and one.

The exponential response $s(\lambda) = e^{-q\lambda}$ is one member of this broader class, but the results in Sections 2.2.1–2.2.4 do not require that specific form.

If the upper support bound is zero, exposure is zero almost surely. Otherwise, assume a finite, strictly positive upper support and the fixed mean in equation (60).

$$0 \leq \Lambda \leq \Lambda_{\max}, \qquad E[\Lambda] = \mu. \tag{60}$$

**Proposition 3. Sharp mean-and-support bounds for convex survival.** For every bounded, non-increasing convex $s$,

$$s(\mu) \le S_s \le \left(1 - \frac{\mu}{\Lambda_{\max}}\right) s(0) + \frac{\mu}{\Lambda_{\max}} s(\Lambda_{\max}). \tag{61}$$

Both endpoints are attainable, so the bounds are sharp.

*Proof.* Jensen's inequality gives

$$E[s(\Lambda)] \ge s(E[\Lambda]) = s(\mu). \tag{62}$$

For every $\lambda \in [0, \Lambda_{\max}]$, convexity places the graph of $s$ below the chord connecting its endpoints:

$$s(\lambda) \le \left(1 - \frac{\lambda}{\Lambda_{\max}}\right) s(0) + \frac{\lambda}{\Lambda_{\max}} s(\Lambda_{\max}). \tag{63}$$

Taking expectations gives the upper bound. The lower endpoint is attained by $\Lambda = \mu$ almost surely, whereas the upper endpoint is attained by

$$P(\Lambda = 0) = 1 - \frac{\mu}{\Lambda_{\max}}, \qquad P(\Lambda = \Lambda_{\max}) = \frac{\mu}{\Lambda_{\max}}. \tag{64}$$

■

Thus, mean and support alone confine survival to sharp endpoints determined by convexity; the exponential form is not required. The upper bound in Proposition 3 corresponds to the classical Edmundson–Madansky bound (Madansky 1959) for the expectation of a convex function with known mean and bounded support; here it is applied to microscopic BNCT exposure to obtain a sharp upper survival bound.

**Proposition 4. Convex-order monotonicity.** Using the standard convex-order notation of Shaked and Shanthikumar (2007), if $\Lambda_A$ and $\Lambda_B$ have the same finite mean and

$$\Lambda_A \le_{\text{cx}} \Lambda_B, \tag{65}$$

then

$$E[s(\Lambda_A)] \le E[s(\Lambda_B)] \tag{66}$$

for every convex survival kernel $s$ for which the expectations exist.

Variance ordering alone is insufficient to imply this convex-order relation.

#### 2.2.2. Treatment ranking and lower-tail constraints

For fixed $\mu$ and $\Lambda_{\max}$, define

$$\ell_s(\mu) = s(\mu) \tag{67}$$

and

$$u_s(\mu) = \left(1 - \frac{\mu}{\Lambda_{\max}}\right) s(0) + \frac{\mu}{\Lambda_{\max}} s(\Lambda_{\max}). \tag{68}$$

**Proposition 5. Sharp survival identified interval.** The complete set of survival values compatible with the specified mean and support is

$$I_s(\mu) = [\ell_s(\mu), u_s(\mu)]. \tag{69}$$

*Proof.* The two endpoints are attained by the extremal distributions in Proposition 3. Any convex mixture of these two distributions retains the same mean and produces the corresponding convex mixture of their survival values. Hence every intermediate value is attainable. ■

The sharp endpoint geometry and the resulting complete identified interval are illustrated in figure 2(b).

Now consider two treatment conditions A and B evaluated under a common survival kernel $s$ and a common upper support bound $\Lambda_{\max}$, with

$$0 < \mu_A < \mu_B \leq \Lambda_{\max}. \tag{70}$$

**Corollary 5. Mean-based treatment-ranking criterion.** Treatment B is guaranteed to satisfy $S_B \leq S_A$ for every compatible pair of microscopic distributions under this common support bound if and only if

$$u_s(\mu_B) \leq \ell_s(\mu_A). \tag{71}$$

If instead

$$u_s(\mu_B) > \ell_s(\mu_A), \tag{72}$$

then mean-based ranking is not identified; compatible distributions can satisfy

$$\mu_B > \mu_A \qquad \text{but} \qquad S_B > S_A. \tag{73}$$

A larger mean exposure therefore does not necessarily imply greater population killing. The result does not reverse the single-target dose-response relation: $s$ is non-increasing. The reversal arises from uncertainty in how exposure is distributed across targets.

Figure 2(c) contrasts the overlapping and separated interval cases.

The same framework yields a direct lower-tail constraint. For any threshold $\lambda_c \geq 0$, define

$$\rho_c = P(\Lambda \leq \lambda_c). \tag{74}$$

Proposition 6 requires only a non-negative, non-increasing survival kernel; convexity is not needed.

**Proposition 6. Cold-tail bound.**

$$S_s \geq \rho_c\, s(\lambda_c). \tag{75}$$

*Proof.* Because $s$ is non-increasing, $s(\Lambda) \geq s(\lambda_c)$ whenever $\Lambda \leq \lambda_c$. Therefore,

$$E[s(\Lambda)] \geq E\left[s(\Lambda)\, \mathbf{1}_{\{\Lambda \leq \lambda_c\}}\right] \geq s(\lambda_c) P(\Lambda \leq \lambda_c). \tag{76}$$

■

Thus, the low-exposure fraction directly constrains survival and captures microscopic coverage information that is absent from the mean alone (figure 2(d)).

Increasing the mean by further enriching already well-exposed targets is therefore not equivalent, under this bound, to reducing the cold tail.

### 2.2.3. High-exposure limits and lower-tail bounds

The next two propositions require only boundedness and monotonicity of $s$; convexity is not needed.

Let $X \geq 0$ describe a fixed relative microscopic exposure pattern and let $a > 0$ scale its overall magnitude:

$$\Lambda_a = aX. \tag{77}$$

Define

$$S_s(a) = E[s(aX)]. \tag{78}$$

Because $s$ is bounded and non-increasing, the limit

$$s_\infty = \lim_{z\to\infty} s(z) \tag{79}$$

exists.

**Proposition 7. High-exposure asymptotic limit.**

$$\lim_{a\to\infty} S_s(a) = s_\infty + [s(0) - s_\infty]P(X = 0). \tag{80}$$

*Proof.* If $X = 0$, then $s(aX) = s(0)$ for every $a$. If $X > 0$, then $aX \to \infty$ and therefore $s(aX) \to s_\infty$. Because $s$ is bounded, dominated convergence gives

$$\lim_{a\to\infty} E[s(aX)] = s(0)P(X = 0) + s_\infty P(X > 0), \tag{81}$$

which is equivalent to the stated expression. ■

For a normalized survival kernel satisfying $s(0) = 1$ and $s_\infty = 0$, this reduces to

$$\lim_{a\to\infty} S_s(a) = P(X = 0). \tag{82}$$

A continuous microscopic distribution need not contain an exact atom at $X = 0$. The lower tail nevertheless remains relevant.

**Proposition 8. Near-zero tail bounds.** For every $\varepsilon > 0$,

$$s(a\varepsilon)P(X \leq \varepsilon) \leq S_s(a) \tag{83}$$

and

$$S_s(a) \leq s(0)P(X \leq \varepsilon) + s(a\varepsilon)P(X > \varepsilon). \tag{84}$$

In particular,

$$S_s(a) \le s(0)P(X \le \varepsilon) + s(a\varepsilon). \tag{85}$$

*Proof.* On $X \le \varepsilon$, $aX \le a\varepsilon$ and monotonicity gives $s(aX) \ge s(a\varepsilon)$, yielding the lower bound.

For the upper bound, split the expectation into $X \le \varepsilon$ and $X > \varepsilon$. On the first event $s(aX) \le s(0)$, whereas on the second $s(aX) \le s(a\varepsilon)$. ■

If $s_\infty = 0$, then for every fixed $\varepsilon > 0$,

$$s(a\varepsilon) \to 0 \tag{86}$$

as $a \to \infty$. Thus, the upper bound separates the probability mass near zero from a term that vanishes at high exposure, including when $P(X = 0) = 0$.

Propositions 7 and 8 describe the contribution of the lower tail when a fixed relative exposure pattern is scaled. For survival kernels satisfying $s_\infty = 0$, the contribution from targets whose relative exposure is bounded away from zero vanishes as the scale increases. The bounds motivate measurement of probability mass near zero, but do not assert that its fraction of total residual survival must tend to one for every bounded, non-increasing kernel. They support experiments that quantify poorly exposed clonogens and examine their biological context.

These asymptotic results identify which part of the microscopic exposure distribution contributes to residual survival; they do not make a claim about the behavior of finite-moment approximation error.

**2.2.4. Propagation to tumor-control probability**

Under the conventional Poisson clonogen approximation,

$$\mathrm{TCP} = e^{-N_0 S}, \tag{87}$$

where $N_0$ is the effective pretreatment clonogen number and $S$ is the surviving fraction.

If the complete survival identified set is the closed interval

$$S_{\min} \le S \le S_{\max}, \tag{88}$$

then its image under the Poisson TCP map is also a closed interval.

**Proposition 9. TCP identified interval.**

$$e^{-N_0 S_{\max}} \le \text{TCP} \le e^{-N_0 S_{\min}}. \tag{89}$$

The ratio of the displayed endpoints is

$$\frac{\text{TCP}_{\max}}{\text{TCP}_{\min}} = e^{N_0 (S_{\max} - S_{\min})}. \tag{90}$$

For two tumors with the same positive effective $N_0$,

$$S_B > S_A \quad \Rightarrow \quad \text{TCP}_B < \text{TCP}_A. \tag{91}$$

Thus, any survival-ranking reversal propagates to TCP ranking within this model. If equation (88) supplies only outer bounds, equation (89) likewise gives TCP outer bounds and equation (90) is only the ratio of those bound endpoints.

This result is conditional on independent Poisson thinning and the specified effective pretreatment clonogen number; alternative dependence structures require a different tumor-control functional.

### 2.2.5. Temporal exposure re-sampling

Temporal averaging depends on whether underexposure is re-sampled across targets or persists within the same targets.

To isolate this mechanism from the trivial effect of adding exposure, consider a fixed-total-exposure thought experiment.

Let $\Theta$ denote a persistent target-specific state. Conditional on $\Theta$, let $X_1, X_2, \ldots$ be non-negative, conditionally independent and identically distributed full-course-equivalent exposure realizations, and assume $E[X_1] < \infty$. For $n$ equal episodes, define

$$\Lambda_n = \frac{1}{n} \sum_{j=1}^{n} X_j \tag{92}$$

and

$$m(\Theta) = E\left[X_j \mid \Theta\right]. \tag{93}$$

Then

$$E[\Lambda_n] = E[m(\Theta)] = \mu \tag{94}$$

for every $n$.

**Proposition 10. Conditional temporal exposure re-sampling theorem.**

$$\Lambda_{n+1} \leq_{\text{cx}} \Lambda_n. \tag{95}$$

Consequently, for every convex survival kernel $s$,

$$E[s(\Lambda_{n+1})] \leq E[s(\Lambda_n)]. \tag{96}$$

Furthermore,

$$\lim_{n\to\infty} E[s(\Lambda_n)] = E\left[s\big(m(\Theta)\big)\right]. \tag{97}$$

*Proof.* Conditional on $\Theta$, define the leave-one-out means

$$A_i = \frac{1}{n}\sum_{j\neq i} X_j\,, \qquad i = 1, \dots, n+1. \tag{98}$$

Their arithmetic mean is $\Lambda_{n+1}$. Hence, for any convex function $\varphi$,

$$\varphi(\Lambda_{n+1}) \leq \frac{1}{n+1}\sum_{i=1}^{n+1} \varphi\,(A_i). \tag{99}$$

Taking conditional expectations, each $A_i$ has the same conditional distribution as $\Lambda_n$, so

$$E[\varphi(\Lambda_{n+1}) \mid \Theta] \leq E[\varphi(\Lambda_n) \mid \Theta]. \tag{100}$$

Averaging over $\Theta$ establishes the convex-order relation. Taking $\varphi = s$ gives the survival ordering.

Conditional on $\Theta$, the strong law of large numbers gives $\Lambda_n \to m(\Theta)$ almost surely. A finite convex kernel is continuous at every positive argument. If the conditional mean is zero, non-negative exposure implies that every exposure realization is zero almost surely; hence all exposure averages are zero on that event. The kernel values therefore converge almost surely to the kernel evaluated at the conditional mean in either case. Since $s$ is bounded, dominated convergence yields the stated limit. ■

Thus, for convex survival kernels, temporal re-sampling reduces the biological consequence of the heterogeneity component that changes between episodes.

With finite second moments,

$$\mathrm{Var}(\Lambda_n) = \mathrm{Var}\big(m(\Theta)\big) + \frac{1}{n} E[\mathrm{Var}(X_1 \mid \Theta)]. \tag{101}$$

This variance decomposition exposes an asymptotic floor. As $n$ increases, the transient component $E[\mathrm{Var}(X_1 \mid \Theta)]/n$ vanishes, whereas the persistent component $\mathrm{Var}\big(m(\Theta)\big)$ remains. Temporal re-sampling can therefore remove only the component of heterogeneity that varies within the same target across episodes; it cannot eliminate heterogeneity encoded in persistent target-specific mean exposure. The model does not specify the biological origin of such persistence, which may in practice reflect stable spatial, transport, or cellular determinants.

If

$$m(\Theta) = \mu \quad \text{almost surely}, \tag{102}$$

repeated re-sampling approaches the homogeneous-response limit $s(\mu)$.

At the opposite extreme, if exposure is perfectly persistent,

$$X_j = m(\Theta) \tag{103}$$

for every episode, then $\Lambda_n$ does not change and no temporal-coverage benefit occurs.

The limiting distinction is therefore between transient within-target variation, which can average out, and persistent target-specific differences, which remain. These two regimes and the resulting variance floor are illustrated in figure 3(a).

### 2.2.6. Event quality and multidimensional microscopic dependence

The preceding results treat microscopic survival as a function of a scalar exposure $\Lambda$. BNCT biology may depend on additional target-specific variables.

To expose dependence on additional target-level context, let

$$Q \in [0,1] \tag{104}$$

denote a target-specific effective event-quality parameter. Conditional on the reaction intensity and event quality, assume that the event count is Poisson with mean equal to that intensity. Conditional also on the count, assume independent, identically distributed event survival contributions whose common mean is one minus the event-quality parameter and does not depend on the event index or count. Repeating the calculation in (41)–(42) with these conditional assumptions gives

$$S = E[e^{-\Lambda Q}]. \tag{105}$$

$Q$ is an abstract latent quantity that may summarize subcellular localization, charged-particle geometry, proximity to sensitive structures, or other determinants of event consequence.

**Proposition 11. Marginal frequency and quality distributions are insufficient.** The marginal distributions of $\Lambda$ and $Q$ do not, in general, determine $E[e^{-\Lambda Q}]$.

A concrete counterexample is obtained from population A:

$$(\Lambda, Q) = (0.5, 0.2) \quad \text{with probability } \tfrac{1}{2}, \tag{106}$$

$$(\Lambda, Q) = (2, 0.8) \quad \text{with probability } \tfrac{1}{2}, \tag{107}$$

and population B:

$$(\Lambda, Q) = (0.5, 0.8) \quad \text{with probability } \tfrac{1}{2}, \tag{108}$$

$$(\Lambda, Q) = (2, 0.2) \quad \text{with probability } \tfrac{1}{2}. \tag{109}$$

The marginals of $\Lambda$ and $Q$ are identical, but

$$S_A = \tfrac{1}{2} e^{-0.1} + \tfrac{1}{2} e^{-1.6} \approx 0.5534, \tag{110}$$

whereas

$$S_B = e^{-0.4} \approx 0.6703. \tag{111}$$

The two populations therefore have identical marginals but different survival; target-level pairing can matter, as illustrated in figure 3(b).

Define the effective event burden

$$Z = \Lambda Q. \tag{112}$$

Then

$$S = E[e^{-Z}]. \tag{113}$$

The full joint distribution of $(\Lambda, Q)$ is not required once the distribution of $Z$ is known. However, the separate marginals $P_\Lambda$ and $P_Q$ do not generally determine $P_Z$.

Its mean satisfies

$$E[Z] = E[\Lambda]E[Q] + \mathrm{Cov}(\Lambda, Q) \tag{114}$$

but this scalar summary does not suffice to identify $E[e^{-Z}]$. Equation (114) also does not provide a universal dependence-ordering criterion. In particular,

$$\frac{\partial^2}{\partial\lambda\,\partial q} e^{-\lambda q} = (\lambda q - 1)e^{-\lambda q} \tag{115}$$

which changes sign across the state space. Thus, the effect of dependence cannot be characterized globally by a fixed supermodular or submodular ordering.

**Proposition 12. Effective-burden stochastic dominance.** If

$$Z_A \geq_{\mathrm{st}} Z_B, \tag{116}$$

where $\geq_{\mathrm{st}}$ denotes first-order stochastic dominance, then

$$E[e^{-Z_A}] \leq E[e^{-Z_B}]. \tag{117}$$

Within this model, first-order stochastic dominance of the effective-burden distribution is therefore sufficient to guarantee lower survival. The stochastic-dominance statement is a direct application of standard first-order stochastic ordering to the decreasing function $e^{-z}$ (Shaked and Shanthikumar 2007).

A corresponding lower-tail bound is

$$E[e^{-Z}] \geq P(Z \leq z_c)e^{-z_c}. \tag{118}$$

Together, Propositions 11 and 12 show that response can depend on the effective-burden distribution even when separate reaction-intensity and event-quality marginals are known.

A corresponding coverage quantity is the fraction of clonogenic targets with small effective burden.

This quantity captures the low-burden tail in the joint frequency-quality representation.

The same problem arises whenever the microscopic state is multidimensional. Let $\Psi$ denote intrinsic target susceptibility and let

$$R = E[g(\Lambda, \Psi)]. \tag{119}$$

If $g$ depends non-additively on $\Lambda$ and $\Psi$, separate summaries of exposure and susceptibility do not, in general, identify their joint biological effect. A biologically adverse dependence occurs when poorly exposed targets also occupy relatively resistant states. Whether such co-occurrence exists is an empirical question; the framework does not assign it to any specific cell class. The same logic applies to mixed-field radiation. If the microscopic response contains biologically relevant cross-component terms, separate component averages cannot generally recover them.

### 2.2.7. Hierarchy of retained information dimensions

Taken together, the preceding results distinguish five information dimensions that can be retained cumulatively to address different sources of ambiguity in BNCT effect prediction. Starting from a mean exposure, adding selected lower-tail descriptors additionally resolves the fraction of poorly exposed targets. Retaining a richer marginal exposure distribution further captures distributional shape beyond the mean and selected tail summaries. Longitudinal measurements add information about whether underexposure is transient or persists within the same targets. Joint measurements of exposure and biological context additionally retain dependence that is invisible to separate marginals.

These information dimensions can narrow an identified response set or improve a validated approximation only insofar as they are relevant to the response functional and admissible microscopic model. The resulting hierarchy is:

mean exposure → selected lower-tail descriptors → richer marginal exposure distribution → longitudinal persistence of underexposure → joint exposure–biological context.

This hierarchy summarizes cumulatively richer retained information, rather than a universal ranking of measurement value, efficiency, or cost (figure 3(c)).

## 3. Discussion

### 3.1. Interpretation and implications

The central contribution of this study is to formalize an information-sufficiency question that sits upstream of conventional BNCT forward modeling. A recent conceptual study organized BNCT effect prediction across observable information, latent cellular boron state, cell-associated capture statistics, microdosimetric context, and biological consequence (Furuya and Fujimura 2026). The present analysis asks what happens when that microscopic description is compressed: does the retained finite representation still determine an expectation-based biological response uniquely? This shifts the emphasis from whether heterogeneity matters, which is already well established, to whether the information needed to propagate heterogeneity into effect has survived coarse-graining. The issue is therefore not whether contemporary BNCT models rely on a mean dose alone, but whether the microscopic inputs required by those models are retained in the available representation or supplied through independently justified assumptions.

That question complements, rather than competes with, increasingly detailed BNCT dosimetry and radiobiology. Photon-isoeffective modeling, stochastic microdosimetric kinetic approaches, cellular-scale PHITS simulations, LISMEC, and recent microdistribution-aware analyses show how specified microscopic information can alter predicted biological effect (González and Santa Cruz 2012; Marcaccio et al 2026; Santa Cruz 2026; Sato et al 2018; Shigehira et al 2023; Shigehira et al 2026). For example, PHITS-based cellular microdosimetry demonstrated sensitivity to intracellular boron localization, whereas LISMEC propagates cellular-scale model parameters into macroscopic BNCT dosimetry (Shigehira et al 2023; Shigehira et al 2026). Such advances sharpen the identifiability problem: before a forward model can support a unique prediction, the inputs on which its response depends must either be retained in the observation or supplied by independently justified assumptions.

For expectation-based endpoints, Theorem 1 gives an exact criterion within the stated distribution-free class. The abstract problem belongs to the broader theory of generalized moment constraints and moment sets (Kemperman 1968; Winkler 1988). The role of Theorem 1 here is not to claim a new general result in moment theory, but to formulate the exact information-sufficiency criterion explicitly for microscopic coarse-graining in BNCT and to connect it to the biological consequences developed below. Finite linear population summaries identify the response if and only if the single-target response kernel lies in the linear span of the constant function and the retained descriptor functions. This is not a claim that coarse summaries are necessarily inaccurate; it specifies when they are sufficient for unique recovery. Because the theorem already permits arbitrary nonlinear post-processing of the retained summaries, computation applied after coarse-graining cannot reconstruct response information that the representation did not retain. Convexity is not part of this criterion; it enters only in the later ordering and bounding results.

Failure of exact identification is therefore a starting point for more informative questions rather than an endpoint. The identified set states which response values remain compatible with the retained information; approximation bounds quantify what can be recovered when the response kernel is close to the span generated by the retained descriptors; and a restricted mechanistic or parametric family can restore identification by adding physical or biological structure. These are distinct routes to useful prediction under incomplete information, and each makes explicit what has been added beyond the original coarse representation.

This distinction is particularly important for mean dose. Mean boron dose and other macroscopic dose quantities remain indispensable for prescription, treatment planning, quality assurance, and comparison (Coderre and Morris 1999; González and Santa Cruz 2012; International Atomic Energy Agency 2001; Sauerwein et al 2025). The theorem addresses the stronger question of whether a mean is exactly sufficient for a target-level response under unrestricted microscopic heterogeneity, even after physical scoring and biological target weighting are matched. Under convex survival assumptions, the mean and an upper support bound can still yield a sharp survival interval, and treatment ranking becomes robust when the compatible intervals are sufficiently separated. A mean can therefore remain highly informative without uniquely determining the population response.

The lower-tail results make this distinction biologically concrete. A non-negligible fraction of targets below an exposure threshold imposes a lower bound on survival, and under the stated high-exposure scaling the contribution from relative exposures bounded away from zero vanishes when the survival kernel tends to zero. These results do not establish a universal clinical dose-escalation plateau. Rather, they identify microscopic coverage—whether poorly exposed targets are actually lifted—as an information dimension that is invisible to the mean alone.

This perspective may be relevant to boron-carrier evaluation. Conventional development appropriately emphasizes tumor accumulation, tumor-to-normal selectivity, retention, and favorable pharmacokinetics (Barth et al 2018; Wittig et al 2008), yet similar mean tumor boron concentrations can arise from different target-level coverage patterns. Cellular-scale microdosimetry has shown that intracellular localization can alter predicted biological effectiveness even when boron quantity is comparable (Shigehira et al 2023), and Nishimura et al (2024) reported greater tumor boron accumulation but lower BNCT efficacy for a carrier with a distinct intracellular localization pattern. These observations do not validate any specific response kernel assumed here, but they illustrate why boron quantity and event-relevant microscopic context should not automatically be treated as interchangeable summaries.

Relatedly, polyarginine-conjugated sodium borocaptate (BSH) has been shown to preferentially accumulate in CD44$^{High}$ tumor-cell populations and to engage defined cell-surface and intracellular targets (Fujimura et al 2020). Such findings motivate target-resolved questions about whether low reaction burden and unfavorable biological state co-occur in the same cells. The framework does not assume that they do; it identifies dependence itself as potentially response-relevant information.

Section 2.2.6 makes the same point in a minimal marked-event model. Survival depends on the distribution of the effective burden $Z = \Lambda Q$, whereas the separate marginal distributions of $\Lambda$ and $Q$ need not determine that burden distribution. More generally, whenever exposure interacts non-additively with susceptibility or other biological context, target-level dependence becomes part of the information required for prediction. Whether poorly exposed targets preferentially occupy resistant states is therefore an empirical question, not an assumption of the framework.

Temporal heterogeneity adds a different form of missing information: target identity over time. Under the fixed-total-exposure re-sampling model, variation that changes within the same target across episodes is progressively averaged out, whereas persistent differences in target-specific mean exposure remain. Thus, a change in tumor-average uptake across treatment sessions does not establish that previously poorly exposed clonogens have been re-sampled into better-exposed states.

Clinical data illustrate why this distinction matters. In a prospective fractionated BNCT study for recurrent head and neck cancer, $^{18}$F-BPA PET was performed before each treatment and tumor-to-normal uptake ratios differed between the two irradiation sessions (Wang et al 2016). These observations demonstrate macroscopic temporal variation, but they cannot determine whether the same microscopic targets remained underexposed. Longitudinal target-resolved measurements would therefore address a different question from repeated measurement of tumor-average uptake. The re-sampling result itself does not imply a clinical benefit from fractionation.

The expectation-based single-target representation also marks a boundary of the present framework. Real tumors contain interacting malignant and non-malignant populations, extracellular structure, vascular constraints, immune components, and spatially coupled responses. Three-dimensional models incorporating patient-derived cancer-associated fibroblasts illustrate how BNCT response can be studied within a multicellular microenvironment (Yamamoto et al 2025). Such systems may require non-local or higher-order response functionals rather than simply additional descriptors within the present single-target theorem.

Taken together, the results suggest evaluating additional measurements by the ambiguity they resolve. Lower-tail descriptors address microscopic coverage, richer marginal distributions retain distributional shape, longitudinal observations address persistence of underexposure, and joint measurements retain dependence between exposure and biological context. Their usefulness depends on the endpoint and admissible microscopic model rather than on a universal hierarchy of measurement value. Empirical studies can test whether a proposed information dimension narrows the identified response set or improves a validated approximation beyond a coarser representation.

Identifiability analysis therefore does not prescribe a measurement modality or a universal dose metric. It constrains the information that a representation must retain for a specified response to be exactly recoverable; biology and physics determine which variables are physically meaningful, measurable, and mechanistically justified. The practical value of the framework is to make explicit whether predictive specificity comes from retained observables, additional assumptions, or an accepted approximation tolerance, rather than attributing it to the coarse summary alone.

### 3.2. Limitations

The analysis is structural and operates in an idealized noiseless population limit. It does not quantify finite-sample uncertainty, imaging or reconstruction error, boron-assay variability, dosimetric uncertainty, or statistical estimation error. Conversely, those uncertainties do not remove the structural ambiguity identified here. The necessity part of Theorem 1 also depends on an admissible distribution class rich enough to contain all finitely supported probability measures; independently justified physiological or mechanistic restrictions can narrow that class and may restore identification from fewer summaries.

Macroscopic dose remains essential for prescription, treatment planning, quality assurance, and comparison (Coderre and Morris 1999; González and Santa Cruz 2012; International Atomic Energy Agency 2001; Sauerwein et al 2025). The results concern stronger biological inferences drawn from coarse summaries. Corollary 5 is formulated for matched microscopic target exposure and does not establish rank reversal between clinical treatment-planning-system plans based on voxel means or dose–volume histograms alone; such an application requires an explicit mapping from planning observables to the clonogen-level exposure distribution. Likewise, the high-exposure bounds do not establish a clinical dose-escalation plateau or identify a resistant phenotype.

The survival results are also conditional on their stated kernel and model assumptions. Convexity is required for the mean-and-support and convex-order results, whereas the lower-tail and high-exposure bounds use non-negative, bounded, non-increasing kernels. The fixed-total-exposure temporal model assumes conditionally independent, identically distributed exposure realizations given a persistent target state and excludes repair, repopulation, evolving pharmacokinetics, normal-tissue toxicity, and dose-rate effects; it is not a recommendation for clinical fractionation. The TCP mapping assumes independent Poisson thinning, and correlated survival or non-Poisson clonogen statistics may require a different tumor-control functional.

The marked-Poisson specialization is deliberately minimal. Its latent variables $\Lambda$, $Q$, $Z$, and $\Theta$ are target-level constructs used to expose information structure, not directly measurable clinical quantities or proposed weighting factors. Track structure, DNA damage complexity, repair, dose rate, boron redistribution during irradiation, mixed-field interaction, and other radiobiological mechanisms may require a richer state description. In particular, the event-quality parameter should not be interpreted as Compound Biological Effectiveness (CBE),

Relative Biological Effectiveness (RBE), or Linear Energy Transfer (LET), and the exponential kernel is not proposed as a universal BNCT survival law.

## 4. Conclusion

For expectation-based BNCT endpoints, exact distribution-free identification from finitely many linear population summaries is possible if and only if the single-target response kernel lies in the linear span of the constant function and the retained descriptor functions, under the stated admissible class. When this condition fails, informative prediction may still be possible through partial identification, controlled approximation, or independently justified restrictions on the microscopic model. The analysis further identifies lower-tail exposure, longitudinal persistence of underexposure, and target-level coupling between exposure and biological context as distinct information dimensions that coarse summaries may fail to retain. The No-Go framework therefore does not define a limit of BNCT itself; it defines the information that must be retained or otherwise supplied before a specified biological response can be uniquely determined.

## Data availability

No new data were created or analyzed in this study.

**Figure Legends**

**Figure 1. Structural identifiability under microscopic coarse-graining in BNCT. (a)** Schematic representation of the multiscale information hierarchy linking observable BNCT information to latent microscopic state and biological response. The microscopic state may include cellular boron distribution, cell-associated capture statistics, and event-relevant context. **(b)** Coarse-graining maps a heterogeneous microscopic population $P$ to a finite vector of retained population summaries, whose components are averages of the retained descriptor functions. **(c)** Distinct microscopic populations can share the same retained summary while producing different biological responses. Thus, equality of the coarse representation does not in general imply equality of the response. **(d)** For an admissible class containing all finitely supported probability measures, Theorem 1 states that exact distribution-free identification holds if and only if the single-target response kernel satisfies $g \in \text{span}\{1, h_1, \ldots, h_m\}$. When this condition fails, there exists at least one attainable retained summary $t$ for which the identified response set is non-singleton. Failure of exact identification does not preclude partial identification, controlled approximation, or prediction under additional justified model restrictions.

**Figure 2. Partial identification and microscopic coverage under unresolved exposure heterogeneity. (a)** Two microscopic exposure distributions with the same positive mean $\mu$ can yield different population survival. Population A is homogeneous at $\Lambda = \mu$, whereas population B places probability $1 - p$ at zero exposure and probability $p$ at $\mu/p$. For the exponential survival specialization, strict convexity gives $S_B > S_A$. **(b)** Sharp survival bounds for a bounded, non-increasing convex single-target survival kernel $s(\lambda)$, given mean exposure $\mu$ and upper support $\Lambda_{\max}$. The lower bound $\ell_s(\mu) = s(\mu)$ is attained by homogeneous exposure, whereas the upper bound $u_s(\mu)$ is given by the chord joining $s(0)$ and $s(\Lambda_{\max})$ and is attained by the corresponding endpoint mixture. All intermediate values are attainable, forming the identified interval $I_s(\mu)$. **(c)** Treatment ranking depends on the relative positions of the compatible survival intervals. When the intervals overlap, mean-based ranking is not identified. When the upper survival bound for treatment B does not exceed the lower bound for treatment A, the ordering $S_B \leq S_A$ is guaranteed for all compatible microscopic distributions under the stated common-kernel and common-support assumptions. **(d)** Lower-tail exposure provides information not contained in the mean alone. For a threshold $\lambda_c$, the fraction $\rho_c = P(\Lambda \leq \lambda_c)$ imposes the survival bound $S_s \geq \rho_c s(\lambda_c)$. The displayed decreasing survival kernel is illustrative; convexity is not required for this lower-tail bound.

**Figure 3. Temporal persistence and joint microscopic context beyond static exposure summaries. (a)** Temporal re-sampling under a fixed-total-exposure model. Rows represent microscopic targets and columns exposure episodes; lighter cells indicate lower full-course-equivalent exposure. In the transient pattern, the identity of poorly exposed targets changes across episodes, whereas in the persistent pattern the same target remains poorly exposed. Bars show the four-episode target-specific averages $\Lambda_4$. The accompanying variance decomposition separates a persistent between-target component, $\mathrm{Var}[m(\Theta)]$, from a transient within-target component that decreases as $1/n$. The curve is an analytic illustration of equation (101), with the dashed line denoting the persistent variance floor. This construction isolates temporal re-sampling and does not represent cumulative dose addition or a recommendation for clinical fractionation. **(b)** Joint dependence between reaction intensity $\Lambda$ and event quality $Q$. Populations A and B have identical marginal distributions of $\Lambda$ and $Q$, but differ in their target-level pairing, as in equations (106)-(111). Under the marked-event specialization, $S = \mathrm{E}[\exp(-\Lambda Q)]$, giving $S_A \approx 0.5534$ and $S_B \approx 0.6703$. Thus, separate marginals need not determine survival because their pairing can matter; this does not imply that every change in pairing changes the response. **(c)** Information dimensions beyond a static mean. The glyphs represent, from left to right, a retained mean, selected lower-tail information, the broader marginal exposure distribution, longitudinal persistence of underexposure, and joint exposure-biological context. The sequence organizes progressively richer information relevant to different sources of ambiguity; it is not a universal ranking of measurement value, efficiency, or cost.

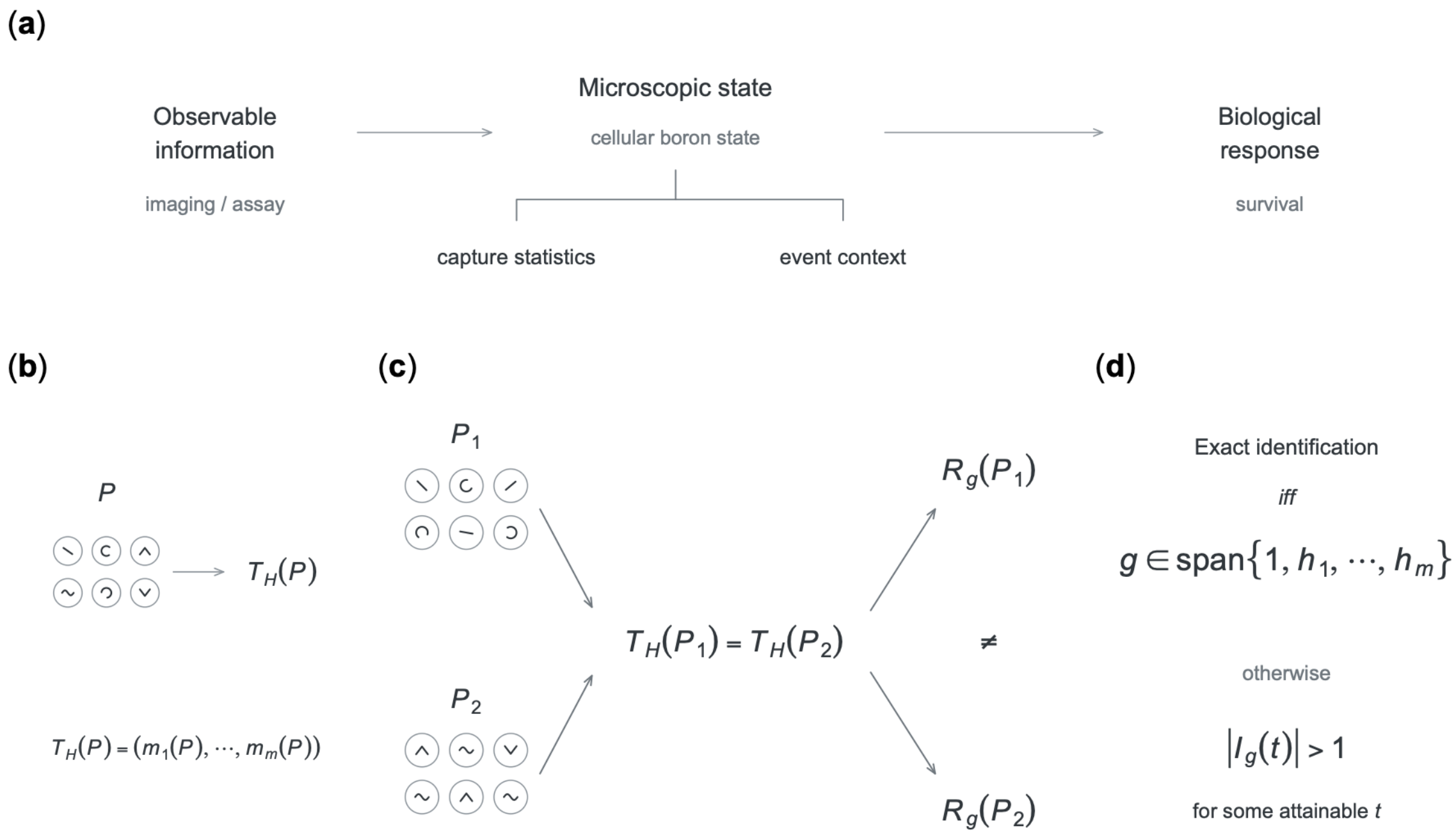
(a)
Observable information
imaging / assay
Microscopic state
cellular boron state
capture statistics
event context
Biological response
survival
(b)
P
T_H(P)
T_H(P) = (m_1(P), ⋯, m_m(P))
(c)
P_1
P_2
T_H(P_1) = T_H(P_2)
R_g(P_1)
≠
R_g(P_2)
(d)
Exact identification
iff
g ∈ span{1, h_1, ⋯, h_m}
otherwise
|I_g(t)| > 1
for some attainable t

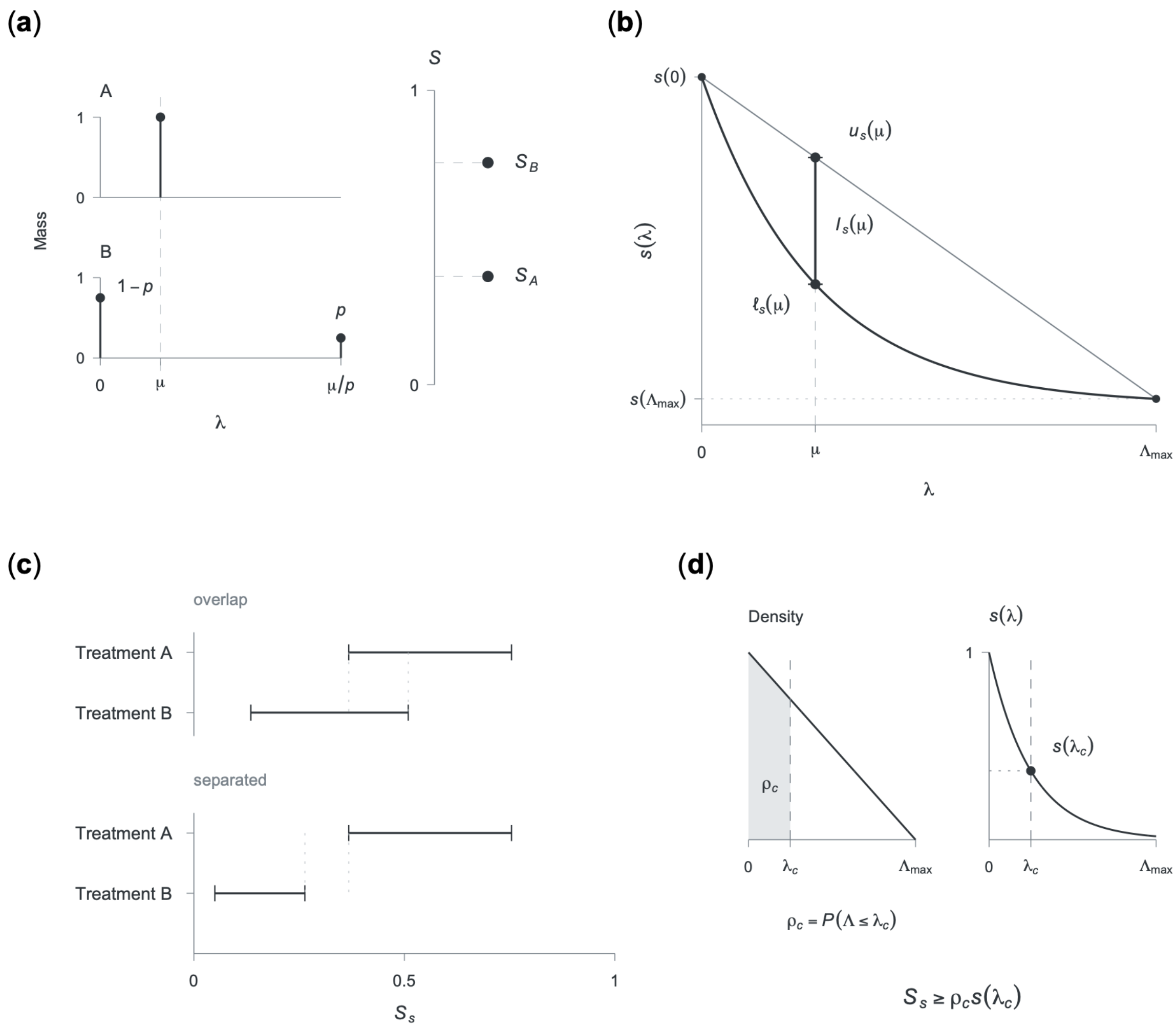
(a)
A
B
Mass
1
0
1 − p
p
μ
μ/p
λ
S
S_B
S_A
(b)
s(0)
u_s(μ)
I_s(μ)
ℓ_s(μ)
s(λ)
s(Λmax)
Λmax
(c)
overlap
Treatment A
Treatment B
separated
0.5
S_s
(d)
Density
ρ_c
λ_c
s(λ_c)
ρ_c = P(Λ ≤ λ_c)
S_s ≥ ρ_c s(λ_c)

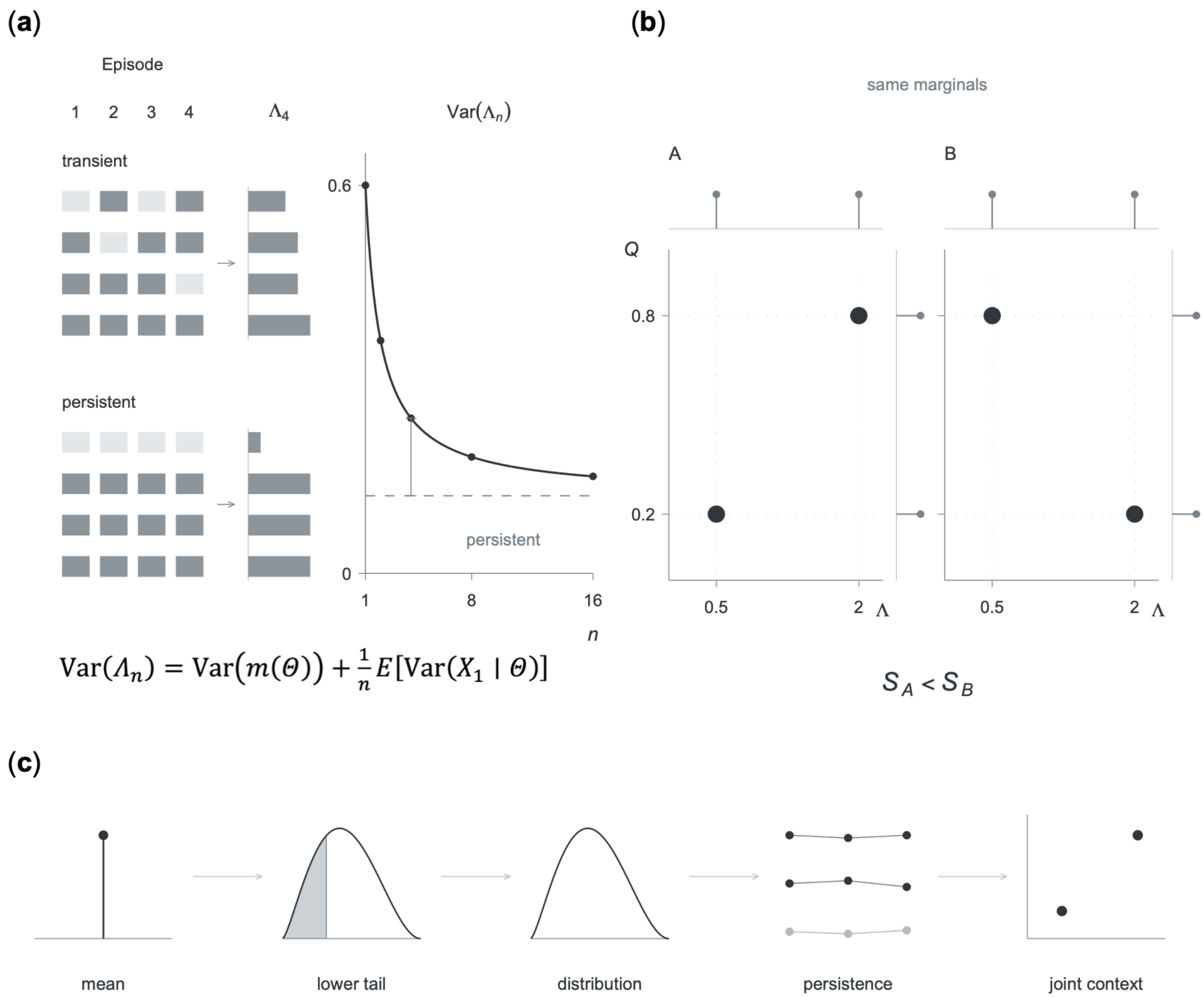
(a)
Episode
1 2 3 4
Λ4
Var(Λn)
transient
persistent
0.6
0
1 8 16
n
persistent
Var(Λn) = Var(m(Θ)) + (1/n) E[Var(X1 | Θ)]
(b)
same marginals
A
B
Q
0.8
0.2
0.5 2 Λ
0.5 2 Λ
SA < SB
(c)
mean
lower tail
distribution
persistence
joint context